\documentclass[authoryear,review,1p,times]{elsarticle}
\usepackage{amsmath,amssymb,indentfirst,epsfig,multirow,amsthm,color,array,amsfonts,graphicx,fullpage,bm,hyperref}
\usepackage[authoryear]{natbib}
\usepackage{amssymb, upgreek}
\usepackage{bm}
\usepackage{arydshln,booktabs}
\usepackage{threeparttable}
\usepackage{framed,color,verbatim}

\usepackage{upgreek,subfigure,url,booktabs}
\usepackage{longtable}
\usepackage{scalefnt}
\usepackage{multirow,url}
\usepackage{threeparttable}
\usepackage{hyperref}
\usepackage{amsmath}
\usepackage{amsfonts}
\usepackage{amssymb}
\usepackage{relsize}
\usepackage{float}
\usepackage{stmaryrd}
\usepackage{caption}
\usepackage{longtable}
\usepackage{framed,verbatim}
\usepackage{lscape}

\usepackage{colortbl}

\usepackage[latin1]{inputenc}
\usepackage[figuresright]{rotating}
\usepackage{url,booktabs}

\usepackage{subfigure}
\usepackage{psfrag}
\usepackage{lineno}

\definecolor{shadecolor}{rgb}{.9, .9, .9}

   {\snugshade\verbatim}%
   {\endverbatim\endsnugshade}

\newdefinition{rmk}{Remark}
\newproof{pf}{Proof}

\hypersetup{
		colorlinks=true,       		% false: boxed links; true: colored links
    	linkcolor=blue,          	% color of internal links
    	citecolor=blue,        		% color of links to bibliography
    	filecolor=magenta,      		% color of file links
		urlcolor=blue,
		bookmarksdepth=4
}

\usepackage{listings}
\usepackage{color}

\definecolor{codegreen}{rgb}{0,0.6,0}
\definecolor{codegray}{rgb}{0.5,0.5,0.5}
\definecolor{codepurple}{rgb}{0.58,0,0.82}
\definecolor{backcolour}{rgb}{0.95,0.95,0.92}

\lstdefinestyle{mystyle}{
    backgroundcolor=\color{backcolour},
    commentstyle=\color{codegreen},
    keywordstyle=\color{magenta},
    numberstyle=\tiny\color{codegray},
    stringstyle=\color{codepurple},
    basicstyle=\footnotesize,
    breakatwhitespace=false,
    breaklines=true,
    captionpos=b,
    keepspaces=true,
    numbers=left,
    numbersep=5pt,
    showspaces=false,
    showstringspaces=false,
    showtabs=true,
    tabsize=2
}

\journal{ArXiv}
\begin{document}
%\linenumbers

\begin{frontmatter}

\title{Zero-Modified Poisson-Lindley distribution with applications in zero-inflated and zero-deflated count data}

\author[label11]{Danillo  Xavier \fnref{label3}}

\author[label11,label13]{Manoel Santos-Neto\fnref{label2}}

\address[label11]{Universidade Federal de Campina Grande\\
Departamento de Estatística, Bairro Universit\'ario, 58429-900, Campina Grande, PB, Brazil
\vspace{.1cm}}

\address[label13]{Universidade Federal de S\~ao Carlos\\
Departamento de Estatística, Rodovia Washington Luis, km 235, 13565-905, S\~ao Carlos, SP, Brazil
\vspace{.1cm}}

 \author[label55]{Marcelo Bourguignon \fnref{label1}}

\address[label55]{Universidade Federal do Rio Grande do Norte\\
 Departamento de Estatística, Lagoa Nova, 59078-970, Natal, RN, Brazil
 \vspace{.1cm}}

\author[label13]{Vera Tomazella \fnref{label4}}

\fntext[label2]{Corresponding author. Email: \texttt{mn.neco@gmail.com}}
\fntext[label3]{Email: \texttt{danilloxavier@gmail.com}}
\fntext[label1]{Email: \texttt{m.p.bourguignon@gmail.com}}
\fntext[label4]{Email: \texttt{vera@ufscar.br}}

\begin{abstract}
%\begin{center}
%{\color{blue} ***************--------------I'm doing an update here!---------------***************}
%\end{center}
The main object of this article is to present an extension of the zero-inflated Poisson-Lindley distribution, called of zero-modified Poisson-Lindley. The additional parameter $\pi$ of the zero-modified Poisson-Lindley has a natural interpretation in terms of either zero-deflated/inflated proportion. Inference is dealt with by using the likelihood approach. In particular the maximum likelihood estimators of the distribution's parameter are compared in small and large samples. We also consider an alternative bias-correction mechanism based on Efron's bootstrap resampling. The model is applied to real data sets and found to perform better than other competing models.

\end{abstract}

\end{frontmatter}

\section{Introduction} ~\label{sec:1}

In many areas of statistical applications like insurance, medicine, ecology and biology, for example, we come across situations where the zeros show up in the count data with a greater or a lesser tendency. Correspondingly, one needs to adapt a count data model by inflating, deflating or truncating the probability associated with a zero count. In this sense many generalizations or modifications have been considered in the literature, see, for example,  \cite{zuur2009}. Also, in \cite{plackett53} the zero-truncated Poisson distribution is presented. A class of zero-modified Poisson models is discussed in \cite{dietzBo00}.  Already, in \cite{borahdeka2001} the Poisson-Lindley (PL) distribution  has been further studied with only inflation of probability at zero.

The Poisson-Lindley (PL) distribution~\citep{san70} is a generalized Poisson distribution \citep{Lindley58,consul89} with probability mass function (pmf) given by

\begin{equation*} \label{eq1}
%\textrm{Pr}(k;\theta)\equiv
\textrm{Pr}(Y=k) = \frac{\theta^2(k+\theta+2)}{(\theta+1)^{k+3}}, \quad \theta > 0, \quad k \in \mathbb{N}.
\end{equation*}

The mean and variance of the PL distribution are given, respectively, by
\begin{equation*}
\textrm{E}(Y) = \frac{\theta+2}{\theta(\theta+1)} \, \, \, \textrm{and} \, \, \,
\textrm{Var}(Y) = \frac{\theta^3+4\theta^2+6\theta+2}{\theta^2(\theta+1)^2}.
\end{equation*}

Also is possible shown that $\textrm{Var}(Y)/\textrm{E}(Y) > 1$, i.e, the PL distribution is over-dispersed, for more details see,~\cite{ghi09}.

Besides, some properties of this Inflated Poisson Lindley (IPL) distribution are discussed and \cite{oluypara07} examine and study relations in zero-adjusted models. Besides, relations for reliability measures in the adjusted and unadjusted models are established and appropriate comparisons including the relative error are presented.  \cite{conc17} present a new family of distributions for count data, the so called zero-modified power series (ZMPS). Also, the Hurdle distribution version of the ZMPS distribution is presented.

In practice, count data are often overdispersed so that alternative distributions such as the Poisson-Lindley may be more appropriate than the Poisson distribution. However, if we are studying count data with excess zeros the PL distribution isn't more adequate. This paper proposes an extension of the PL distribution called of zero-modified Poisson-Lindley (ZMPL). This extension takes into account the inclusion of an additional parameter $\pi$ which has a natural interpretation in terms of either zero-inflated or zero-deflated proportion.

It is well-known that the MLEs are  widely used to estimate the unknown parameters of the probability distributions due to their various desirable properties; for example, the MLEs are asymptotically  unbiased, consistent, and asymptotically normal. However, many of these properties depend on an extremely large sample sizes. Those properties, such as unbiasedness,
may not be valid for small or even moderate sample sizes, which are more practical in  real data applications. Therefore, some bias-corrected techniques for the MLEs are desired in practice, especially when the sample size is small.  In this paper, we consider an alternative bias-corrected maximum likelihood (ML) estimators based on Efron's bootstrap resampling. Indeed, such a bias-correction technique has been applied successfully for parameter estimation in other distributions and models; see, for example, \cite{Efron:1986aa}.

The rest of this article is organized as follows. In Section \ref{sec2}, we present the ZMPL distribution and some of its properties.  Section \ref{sec3} we briefly discuss point and interval estimation by the ML method and of their corrected version for the ZMPL distribution. Numerical results from Monte Carlo simulation experiments are presented and discussed in Section~\ref{sec4}. Section \ref{sec5} two well-known data sets are considered for an empirical comparison. The paper ends with a discussion of extensions and other contexts with the ideas could be implemented in Section~\ref{sec6}.

\section{Zero-Modified Poisson-Lindley (ZMPL) distribution }
\label{sec2}

%\begin{center}
%{\color{blue} ***************--------------I'm doing an update here!---------------***************}
%\end{center}

A random variable $X$ is said to have the ZMPL distribution if its probability mass function is given by
\begin{eqnarray}\label{pdf}
%\textrm{Pr}(k;\theta,\pi)\equiv
 \textrm{Pr}(X=k)=\left\{\begin{array}{ll}
\pi + (1 - \pi)\frac{\theta^2(\theta + 2)}{(\theta + 1)^3},&k = 0,  \\
&\\
(1 - \pi)\frac{\theta^2(k + \theta + 2)}{(\theta + 1)^{k+3}},&k \in \mathbb{N}^{ {\tiny +}}. \\
\end{array}\right.
\end{eqnarray}

We will denote this distribution as ZMPL$(\theta, \pi)$.
For the parameter $\pi$ it is presupposed that  $-\frac{\theta^2(\theta + 2)}{\theta^2 + 3\,\theta + 1} \leq \pi \leq 1$. The parameter $\pi$ is called {\it zero-modification parameter} and different values lead to different modifications of the ZMPL distribution:
\begin{itemize}

\item If $\pi = -\frac{\theta^2(\theta + 2)}{\theta^2 + 3\,\theta + 1}$, then the distribution (\ref{pdf}) becomes the zero-truncated Poisson-Lindley distribution \cite{ghi08}, where the parameter $\pi$ cancels out and no longer appears as a model parameter, i.e., there is no chance at all of getting a zero observation into the sample;

\item  For $\pi \in \left(-\frac{\theta^2(\theta + 2)}{\theta^2 + 3\,\theta + 1}, 0\right)$, this yields a zero-deflated Poisson-Lindley distribution. That is, less zeros occur, than expected under the Poisson-Lindley distribution. Such models are denoted as zero-deflated Poisson-Lindley distribution. Zero-deflated rarely arise in practice;

\item If $\pi = 0$, than the corresponding ZMPL distribution is the usual Poisson-Lindley distribution \citep{san70};

\item For $\pi \in (0, 1)$, this yields a zero-inflated Poisson-Lindley distribution, which is a Poisson-Lindley distribution
with a proportion of additional zeros \citep{borahdeka2001};

\item If $\pi = 1$, than the corresponding zero-modified distribution is the degenerated at zero one.

\end{itemize}

%\subsubsection{Characterization}

%In this section we pres Some of these results, generally, aren't presented for a discrete model.

%Using the following relation
%\begin{equation*}\label{relat}
%\sum_{m=0}^k \frac{m+\theta+2}{(\theta+1)^{m+3}} = \frac{(\theta+1)^{k+3} -  (\theta^2 +(k+3)\theta +1)   }{\theta^2(\theta+1)^{k+3}}.
%\end{equation*}

The corresponding cumulative distribution function (c.d.f.) is given by

\begin{equation*}\label{cdf}
\textrm{Pr}(X \leq k) = \left\{
\begin{array}{lcc}
0, & \mbox{if} & k < 0;\\
\pi + (1-\pi)\left\{1 - \frac{[\theta^2+ ([k]+3)\theta +1] }{(\theta+1)^{[k]+3}} \right\}, &\mbox{if} & k \geq 0,
\end{array}
\right.
\end{equation*}
where $[k]$ is the integer part of $k$. The survival function of $X$, when $\frac{1}{(\theta+1)}<1$, is
\begin{equation*}\label{cdf}
\textrm{Pr}(X \geq k) = \sum_{m=[k]}^\infty \textrm{Pr}(m;\theta,\pi) = \left\{
\begin{array}{lcc}
1, & \mbox{if} & k \leq 0;\\
(1-\pi)\left\{ \frac{(\theta+1)^2 + [k]\theta}{(\theta+1)^{[k]+2}}  \right\}, &\mbox{if} & k > 0.
\end{array}
\right.
\end{equation*}

Let $X$ denotes a random variable with probability mass function given in \eqref{pdf}. The quantile function, say $\textrm{Q}_X(p)$, defined by $F_X^{-1}(p)$ is given by

\begin{eqnarray*}\label{qf}
\textrm{Q}_X(p;\theta,\pi)=\left\{\begin{array}{ll}
0,& \text{if} \quad p_\pi < 0,  \\
&\\
\textrm{Q}_Y(p_\pi; \theta),& \text{if} \quad p_\pi \geq 0, \\
\end{array}\right.
\end{eqnarray*}
where $p_\pi =  (p-\pi)/(1-\pi)$  and $\textrm{Q}_Y(p_\pi; \theta)$ is the quantile function of a Poisson-Lindley distribution. \cite{borahdeka2001} present some properties of the ZMPL distribution.

Fisher index of dispersion~\citep[][p. 163]{johnson2005}  of the random variable $X$ , defined by $\textrm{FI}(X) = \textrm{Var}(X)/\textrm{E}(X)$ is given by
%A commonly used variability measure of a random variable $X$ is the Fisher index of dispersion, defined by $\textrm{FI}(X) = \textrm{Var}(X)/\textrm{E}(X)$ is a measure of aggregation or disaggregation, for more details see~\cite{johnson2005}, pag. 163.
%The FI of the ZMPL distribution is given by
\begin{equation*}\label{idy}
\textrm{FI}(X) = \left\{ \frac{\textrm{E}(X^2)}{\textrm{E}(X)} -  \textrm{E}(X)\right\} = \pi \, \mu_{PL}  + \frac{(\theta^3 + 4\theta^2+6\theta + 2)}{\theta(\theta+1)(\theta+2)} =  \pi \, \mu_{PL} +  \textrm{FI}(Y),
\end{equation*}
where $\mu_{PL}$ and  \textrm{FI}(Y) are, respectively, the mean and the Fisher index of dispersion of a Poisson-Lindley distribution. Thus, the ZMPL distribution presents underdispersion when
$\theta > \sqrt{2}$ and $\pi \in \left[-\frac{\theta^2(\theta + 2)}{\theta^2 + 3\,\theta + 1}, 0\right)$; and overdispersion when $\pi \in [0, 1)$.
% Figure~\ref{fig:fi} we show the form of the Fisher index of dispersion when varying $\theta$ and $\pi$. Note in Figure~\ref{fig:fia} that the o ZMPL model is overdispersed.  Already, in Figure~\ref{ig:fib} the ZMPL model can be over- or under-dispersed.
%
%\begin{figure}[!htb]
%\centering
%\subfigure[\label{fig:fia}$0 \leq\pi \leq 1$]{\includegraphics[width=7.5cm,height=7.5cm]{3d.pdf}}
%\subfigure[\label{ig:fib}$\pi \leq 0 $]{\includegraphics[width=7.5cm,height=7.5cm]{3d2.pdf}}
%\caption{Plot of the Fisher index of dispersion of ZMPL distribution.}
%\label{fig:fi}
%\end{figure}

%\begin{Theorem}
%The p.m.f. given in \eqref{pdf} can be written as a Hundle distribution by
%\[
%\textrm{Pr}(X=k) = (1-\omega) \, I_{\{0\}}(k) + \omega\, \textrm{Pr}(T=k), \quad k \in \mathbb{N},
%\]
%
%where $T$ is a random variable with zero-truncated Poisson-Lindley distribution and $\omega = (1-\pi) \cdot [1-\textrm{Pr}(Y=0)]$ with $\omega \in [0,1]$.

%\end{Theorem}

\subsection{R package}

The theoretical results has been implemented into a piece of statistical software: the {\tt zmpl} package for {\tt R}~\citep{r2017}.
To install this package, the {\tt R} code below must be used.

\begin{lstlisting}[language=R]
devtools::install_github("zmpldistribution/zmpl")
\end{lstlisting}

This package contain a collection of utilities for analyzing data from ZMPL distributions. Some of the functions are: {\tt dzmpl()}, {\tt pzmpl()}, {\tt qzmpl()}, {\tt rzmpl()}, {\tt mle()}, {\tt fi.zmpl()} and {\tt grad.test()}.

\section{Inference}\label{sec3}

This section is concerned with the estimation of the two parameters of interest. We
consider two estimation methods, namely, moments and maximum likelihood.

\subsection{Method of moments estimator}

Let $x_1, x_2, \ldots, x_n$ be a random sample of size $n$ from the ZMPL distribution
with probability mass as given in (\ref{pdf}). The sample arithmetic and quadratic means are
defined by
$$\bar x = \frac{1}{n}\sum\limits_{i=1}^{n}x_i \quad \textrm{and} \quad s_{x^2}= \frac{1}{n}\sum\limits_{i=1}^{n}x_i^{2},$$
respectively.

We have that the first two moments of $X$ are given by
$\textrm{E}(X) = (1 - \pi)(\theta + 2)/[\theta(\theta + 1)]$ and $\textrm{E}(X^2) = (1 - \pi)[(\theta + 2)^2 + 2]/[\theta^2(\theta+1)]$. We can use the arithmetic and quadratic means such that
$$\textrm{E}(X) = \bar x \quad \textrm{and} \quad \textrm{E}(X^2) = s_{x^2},$$
and then solve them for $\theta$ and $\pi$, the moment estimators of $\theta$ and $\pi$ are defined as
\begin{equation*}\label{momentsEst}
\widetilde{\theta} = \displaystyle \frac{\left\{(2\bar x  - s_{x^2}) +  \displaystyle \sqrt{(s_{x^2} - 2\bar x  + 2 \bar x s_{x^2}})\right\}}{ \displaystyle (s_{x^2}-\bar x)} \quad \textrm{and} \quad \widetilde{\pi} = 1 - \frac{\widetilde{\theta}(\widetilde{\theta}+1)\bar x}{\widetilde{\theta} + 2},
\end{equation*}
where $\bar x = s_{x^2}$, if and only if $x_i = 0$ or $1$ for all $i = 1, 2, \ldots n$. A data set where all observations are zeros and ones is not worth analyzing for this distribution. This situation, of course, will not lead to any estimate of $\theta$. However, such situation may arise in a simulation experiment when $n$ and/or $\theta$ are very small or $\pi$ is great. For this reason, we will assume throughout this paper that $\bar x \neq s_{x^2}$.

 Note that, $s_{x^2} = s^2 - \bar x^2$ and thus we can written the moment estimator of $\theta$ as follows
 \[
 \widetilde{\theta} = \displaystyle \frac{\left\{[2\bar x - (s^2 - \bar x^2) ] +  \displaystyle \sqrt{(s^2 - \bar x^2)(1 + 2\bar x) - 2\bar x }\right\}}{ \displaystyle (s^2 - \bar x^2-\bar x)},
 \]
 where $s^2 =  (1/n)\sum_{i=1}^{n}(x_i-\bar x)^{2}$. An iterative method for finding the maximum likelihood (ML) estimators has to be employed. To start the iterative procedure, we could use the initial value obtained by the method of moments.

\subsection{Maximum likelihood estimators}

Let {\color{blue}$x_1, x_2, \ldots, x_n$} be a random sample from ZMPL distribution distribution. Then, the corresponding likelihood function for ${\bm \lambda}= (\theta,\pi)^\top$ is given by
\begin{eqnarray}
\textrm{L}({\bm \lambda};{\bm x}) &=&  \displaystyle\prod_{i=1}^{n}\left\{\pi + (1-\pi)\frac{\theta^2(\theta+2)}{(\theta+1)^3} \right\}^{\textrm{I}_{\{0\}}(x_i)}
 \left\{(1-\pi)\frac{\theta^2(x_i + \theta+2)}{(\theta+1)^{x_i+3}} \right\}^{1-\textrm{I}_{\{0\}}(x_i)} \nonumber \\
 &=& \left\{\pi + (1-\pi)\frac{\theta^2(\theta+2)}{(\theta+1)^3} \right\}^{\sum_{i=1}^n \textrm{I}_{\{0\}}(x_i)}  \displaystyle\prod_{i=1}^{n} \left\{(1-\pi)\frac{\theta^2(x_i + \theta+2)}{(\theta+1)^{x_i+3}} \right\}^{1-\textrm{I}_{\{0\}}(x_i)} \label{prod} \\
 &=& \left\{\pi + (1-\pi)\frac{\theta^2(\theta+2)}{(\theta+1)^3} \right\}^{n_0} \left\{\frac{(1-\pi)\theta^2}{(\theta+1)^3}\right\}^{n-n_0} \displaystyle\prod_{i=1}^{n} \left\{\frac{(x_i + \theta+2)}{(\theta+1)^{x_i}} \right\}^{1-\textrm{I}_{\{0\}}\, x_i},  \nonumber
\end{eqnarray}
where $n_0 = \sum_{i=1}^n \textrm{I}_{\{0\}}(x_i)$ is the number of zeros in sample and $\textrm{I}_A(\cdot)$
is once again the indicator function of the set $A$. Hence, the respective log-likelihood obtained from \eqref{prod} can be expressed as
\begin{eqnarray}
\ell({\bm \lambda};{\color{blue}{\bm x}})  &=& n_0 \log \left\{\pi + (1-\pi)\frac{\theta^2(\theta+2)}{(\theta+1)^3} \right\} + (n-n_0)\log \left\{\frac{(1-\pi)\theta^2}{(\theta+1)^3}\right\} + \sum_{i=1}^{n} [1-\textrm{I}_{\{0\}}(x_i)] \log(x_i+\theta + 2)\nonumber \\
&& - \log(\theta+1)  \sum_{i=1}^{n} [1-\textrm{I}_{\{0\}}(x_i)]\, x_i.
\label{like}
\end{eqnarray}

We obtain the score by taking derivates of the corresponding log-likelihood function with respect to the unknown parameters as
\begin{eqnarray} \label{score}
\ell_{\pi}^{'} &=& \frac{n_0\left(1-\frac{(\theta+2)\theta^2}{(\theta+1)^3}\right)}{\pi+(1-\pi)\frac{(\theta+2)\theta^2}{(\theta+1)^3}} - \frac{n-n_0}{(1-\pi)}, \nonumber \\
\ell_{\theta}^{'} &=& \frac{n_0 (1-\pi)\left( \frac{(3\theta^2+4\theta)}{(\theta+1)^3}   -\frac{3(\theta+2)\theta^2}{(\theta+1)^4}\right)}{\pi+(1-\pi)\frac{(\theta+2)\theta^2}{(\theta+1)^3}} - (n-n_0) \frac{(\theta-2)}{\theta(\theta+1)}  \nonumber \\
&& + \sum_{i=1}^{n} \frac{[1-\textrm{I}_{\{0\}}(x_i)]}{(x_i+\theta+2)} - \frac{1}{(\theta+1)}  \sum_{i=1}^{n} [1-\textrm{I}_{\{0\}}(x_i)] \,x_i.
\end{eqnarray}

From \eqref{score}, the maximum likelihood of $\pi$ is  $\hat \pi = \left[1-\left(1-\frac{n_0}{n}\right)\frac{(\hat \theta+1)^3}{(\hat \theta^2+3\hat \theta+1)}\right]$. Because the solution
of the equation $\ell_{\theta}^{'}=0$ obtained in \eqref{like} doesn't have a closed-form, we maximize the log-likelihood function given in \eqref{like} on $\theta$ by using a non-linear
optimization algorithm for determining the maximum likelihood estimates of $\theta$.

Now, we obtain the respective expected Fisher information matrix by taking derivatives of the
elements of the score vector given in \eqref{score} with respect to the unknown parameters as
\begin{equation}
i({\bm \lambda}) = \begin{bmatrix}
 i_{\theta \theta}&  i_{\theta \pi}\\
 i_{\pi \theta} &  i_{\pi \pi}
\end{bmatrix},
\label{fisher}
\end{equation}
with
\begin{eqnarray*}
i_{\pi \pi}&=&n \left\{ \frac{ \left[ 1-{\frac {{\theta}^{2} \left( \theta+2 \right) }{ \left(
\theta+1 \right) ^{3}}} \right] ^{2}} {\left[ \pi+{\frac { \left( 1-\pi
 \right) {\theta}^{2} \left( \theta+2 \right) }{ \left( \theta+1
 \right) ^{3}}} \right]} + \frac{\left[ 1- \left( \pi+{\frac { \left( 1
-\pi \right) {\theta}^{2} \left( \theta+2 \right) }{ \left( \theta+1
 \right) ^{3}}} \right)  \right]} { \left( 1-\pi \right) ^{2}} \right\}, \\
 i_{\theta \pi}=i_{\pi \theta}&=&n \left[ \frac {(3\theta^2-4\theta)}{ \left( \theta+1 \right) ^{3}}
-{\frac {3{\theta}^{2} \left( \theta+2 \right) }{ \left( \theta+1
 \right) ^{4}}} \right] +
 \frac{n \left[ 1-{\frac {{\theta}^{2} \left( \theta
+2 \right) }{ \left( \theta+1 \right) ^{3}}} \right]
 \left[ {
\frac { \left( 1-\pi \right)  \left( 3\theta^2+4\theta \right) }{
 \left( \theta+1 \right) ^{3}}}-{\frac{3 \left( 1-\pi
 \right) {\theta}^{2} \left( \theta+2 \right) }{ \left( \theta+1
 \right) ^{4}}} \right]}{\left[ \pi+{\frac { \left( 1-\pi \right) {
\theta}^{2} \left( \theta+2 \right) }{ \left( \theta+1 \right) ^{3}}}
 \right]},  \\
i_{\theta \theta}&=&\frac{n(1-\pi)^2 \left[{\frac{3\theta^2+4\theta}{ \left( \theta+1 \right) ^{3}}}-{\frac {3 {\theta}^{2} \left( \theta+2 \right) }{ \left( \theta+1
 \right) ^{4}}} \right] ^{2}}{ \left( \pi+{\frac { \left( 1-\pi \right)
{\theta}^{2} \left( \theta+2 \right) }{ \left( \theta+1 \right) ^{3}}}
 \right)}%\\
+ n\left\{ \left[ \frac{3\theta^3-7\theta^2+4\theta+2}{\theta^2(\theta+1)^2} \right]  \left[ 1- \left[ \pi+{\frac { \left( 1-\pi \right) {\theta}^{2} \left( \theta+2 \right) }{
 \left( \theta+1 \right) ^{3}}} \right]  \right] \right\}
\\
&& - n(1-\pi)\left\{\frac{(\theta^7+ 7\theta^6+ 19\theta^5+ 23\theta^4+10\theta^3-12\theta^2-16\theta+8)}{(\theta+2)(\theta+1)^5} -\frac{\theta^2}{(\theta+1)} \Phi\left(\frac{1}{\theta+1};1;\theta\right)   \right\},
\end{eqnarray*}
where $\Phi\left(\frac{1}{\theta+1};1;\theta\right)$ is the Lerch zeta-function. An integral representation is given by $\Phi\left(\frac{1}{\theta+1};1;\theta\right) = (\theta+1) \int_{0}^{1} \frac{u^{\theta+1}}{(\theta+1-u)}{\rm d}u$.

\subsection{Bias-corrected estimators}

A methodology for bias-correcting estimators is by bootrstrap resampling~\citep{Efron:1979aa}.  Let be a random sample ${\bm x} = (x_1,x_2,\ldots,x_n)^{\top}$ from the random variable $X$ with common distribution $\mathcal{D}$. Define $\vartheta = g(\mathcal{D})$ a function of $\mathcal{D}$ known as parameter and consider $\ddot \vartheta = s({\bm x})$ an estimator of $\vartheta$. 
%In the parametric form of the bootstrap, we obtain, from the original sample ${\bm x}$, a large number of pseudo-sample ${\bm x}^{*}=(x_1^{*},x_2^{*},\ldots,x_n^{*})^{\top}$ , obtaining the corresponding bootstrap replicates of $\ddot \vartheta$, $\ddot \vartheta^{*} = s({\bm x}^{*})$, and using the empirical distribution of $\ddot \vartheta^{*}$, estimate the distribution function of $\ddot \vartheta$.   
We consider that $\mathcal{D}$ belongs to a parametric family which is known and has finite dimension, $ \mathcal{D}_{\nu}$. 
Using a consistent estimator for $\nu (\mathcal{D}_{\ddot \nu})$ we can obtain a parametric estimate for $\mathcal{D}$. Therefore, we can written the bias of the estimator $\ddot \vartheta = s({\bm x})$ as $B_{\mathcal{D}}(\ddot \vartheta, \vartheta) = \textrm{E}_{\mathcal{D}}(s({\bm x})) - g(\mathcal{D})$. Then the  bootstrap bias estimate can be expressed as
$$B_{  \mathcal{D}_{\ddot \nu} }(\ddot \vartheta, \vartheta) = \textrm{E}_{  \mathcal{D}_{\ddot \nu} }(s({\bm x})) - g( \mathcal{D}_{\ddot \nu} ).$$

Finally, the bootstrap bias estimate, calculated from the $B$ replicates of $\ddot \vartheta$, is $B_{  \mathcal{D}_{\ddot \nu} }(\ddot \vartheta, \vartheta) = \bar {\ddot \vartheta}^* -  s({\bm x})$. Then a second order bias-corrected estimator is givem by
\[
\hat \vartheta_{bc} = s({\bm x}) - B_{  \mathcal{D}_{\ddot \nu} }(\ddot \vartheta, \vartheta)  = 2\ddot \vartheta - \bar {\ddot \vartheta}^*.
\]
where $\bar {\ddot \vartheta}^* = \frac{1}{B} \sum_{b=1}^{B} \ddot \vartheta_b^* $ is a approximate of approximate the expected value $\textrm{E}_{  \mathcal{D}_{\ddot \nu} }(s({\bm x})) $.

%
%By generating $B$ bootstrap sample $({\bm x}_1^*,{\bm x}_2^*,\ldots,{\bm x}_B^*)$, independently, and obtaining the corresponding bootstrap replicates $(\ddot \vartheta_1^*,\ddot \vartheta_2^*,\ldots,\ddot \vartheta_B^*)$ one can  by 
\subsection{Confidence Interval {\color{blue}(CI)}}

\subsubsection{Asymptotic Confidence Interval (aCI) }

Under some regularity conditions, $\hat{\bm \lambda}$ is a consistent estimator of ${\bm \lambda}$ and it has a distribution that is asymptotically normal. The, $\sqrt{n}(\hat{\bm \lambda} - {\bm \lambda}) \to \mathcal{N}_2({\bm 0}, j({\bm \lambda})^{-1})$, as $n\to \infty$, where $ j({\bm \lambda}) = \lim_{n\to \infty }\frac{1}{n}i({\bm \lambda} )$, with $i({\bm \lambda} )$ being
the expected Fisher information matrix in \eqref{fisher}  and $\to$ denotes convergence in
distribution to. Note that $i({\bm \lambda} )^{-1}$ is a consistent estimator of the asymptotic variance-covariance matrix of ${\bm \lambda}$. In practice, one may approximate the expected Fisher information matrix by its observed version, whereas the elements of the diagonal of the inverse of this matrix can be used to approximate the corresponding standard errors \citep[see][for details about the use of observed versus expected Fisher information matrices]{Efron:1978aa}.

As a result above, the asymptotic $100(1-\alpha)\%$ confidence intervals (aCI) for $\theta$ and $\pi$
are given by, respectively

$$
\hat \theta \pm z_{\alpha/2} \widehat{\mbox{se}(\hat \theta)} \quad \mbox{and} \quad \hat \pi \pm z_{\alpha/2} \widehat{\mbox{se}(\hat \pi)},
$$
where $\widehat{\mbox{se}(\hat \theta)}$ and $\widehat{\mbox{se}(\hat \pi)}$ are the estimated
asymptotic standard error (se) of the maximum likelihood estimador of $\theta$ and $\pi$, respectively, and $z_{\alpha/2}$ is the $(\alpha/2)$th quantile of the standard normal distribution.

%From Theorem~\ref{teo2}, we obtain the following asymptotic $100(1-\alpha)\%$ confidence intervals for $\theta$ and $\pi$, respectively
%$$
%\tilde \theta \pm z_{\alpha/2} \widetilde{\mbox{se}(\tilde \theta)} \quad \mbox{and} \quad \tilde \pi \pm z_{\alpha/2} \widetilde{\mbox{se}(\tilde \pi)},
%$$
%%
%where $\widetilde{\mbox{se}(\tilde \theta)}$ and $\widetilde{\mbox{se}(\tilde\pi)}$ are the estimated
%asymptotic standard error (se) of the moment estimador of $\theta$ and $\pi$, respectively, and $z_{\alpha/2}$ is the $(\alpha/2)$th quantile of the standard normal distribution.

\subsubsection{Percentile Confidence Intervals (pCI)}

With the empirical distribution $\mathcal{ \ddot D}$ of $\ddot \vartheta$ obtaining
 by bootstrap, one can construct percentile confidence intervals (pCI), with approximate coverage $1-\alpha$, $0<\alpha<1/2$, by computing, by computing the percentiles $\alpha/2$ and $1-\alpha/2$ of $ \mathcal{ \ddot D}$.
 The pCI is given by
 \[
 [\mathcal{ \ddot D}^{-1}(\alpha/2); \mathcal{ \ddot D}^{-1}(1-\alpha/2)].
 \]

 After arranging in increasing order the $B$ bootstrap replicates of $\ddot \vartheta$, $\ddot \vartheta_b^* = s({\bm x}_b^*)$ we calculate the lower and upper limits of the percentiles interval as the integer parts of $B \cdot (\alpha/2)$ and $B \cdot (1- \alpha/2)$, respectively. 

\subsection{Hypothesis test}

Consider the hypotheses  $\mathcal{H}_0: \pi = 0$ and $\mathcal{H}_1: \pi \neq 0$. The interest lies in testing the null hypothesis (PL distribution) against the alternative  hypothesis (ZMPL distribution).

\subsubsection{Gradient test}

Let ${\bm x} = (x_1,x_2,\ldots,x_n)^{\top}$ be a random sample of size $n$ from the ZMPL distribution, each $x_i, i=1,2,\ldots,n$, having p.m.f. \eqref{pdf}. The gradient statistic, $S_g$, is given by

\begin{equation*}
S_g =  n \, \hat \pi^2\, (\hat \theta^2 + 3\hat \theta +1),
\label{grad}
\end{equation*}
where $\hat {\bm \lambda}$ is unrestricted maximum likelihood estimator of ${\bm \lambda}$. Asymptotically,
$S_g$ has a central chi-square distribution with one degree of freedom under $\mathcal{H}_0$.

\section{Simulation}\label{sec4}
%\begin{center}
%{\color{blue} ***************--------------I'm doing an update here!---------------***************}
%\end{center}
%
%In other words, pseudo-random numbers from the two-parameter Poisson-Lindley distribution can be obtained use the following R script:

In this section, we conduct a study based on Monte Carlo (MC) simulations to assess the performance of the ML estimators and of their corrected version. The simulations
were conducted using {\tt R} language \citep[][]{r2017}. The ML estimators of the parameters $\theta$ and
$\pi$ were obtained by maximizing  the log-likelihood function using the BFGS method by package {\tt maxLik} \citep[][]{Henningsen:2011aa}. For each MC replication and for each ML estimate of the parameters of the model, we obtained interval estimates of the asymptotic type, of the percentile type (bootstrap). All intervals were obtained by estimating the two limits independently. The scenario of this simulation study considers 5 000 MC replications and 1 000 bootstrap replications in each case, sample sizes $n =  35, 60, 90$ and $120$.  The values of $\theta$ and $\pi$ were fixed at $\theta \in \{1.5, 2.0\}$ and $\pi \in \{ -0.10, 0.00, 0.10\}$.

The evaluation of point estimation was conducted based on the following measures for each sample size: the mean of estimates (omitted), bias, variance (omitted) and mean squared error. In what concerns interval estimation, we display the means of the empirical coverage probabilities, obtained from the relative frequencies of which the true parameter value belongs to the intervals. The observed frequencies at which the lower limit of the interval was larger (smaller) than the true parameter values are also indicated.

Table~\ref{tabaplic} shows the results of numerical evaluation of point estimators of the parameters of ZMPL distribution. Note that the usual ML estimators of the parameters $\theta$ and $\pi$
are considerably more biased than their corrected versions via bootstrap. For $n=60$, $\pi=-0.10$ and $\theta=1.5$, we noted bias for $\hat {\bm \lambda}$ and $\hat {\bm \lambda}_{bc}$ equal to $(0.090, -0.042)$ and $(0.005, 0.018)$, respectively. That is,  the uncorrected estimators $\hat \theta$ and $\hat\pi$  are, respectively, about 18 and 2 times more biased than the proposed corrected estimators.  Moreover, by the asymptotic properties of the ML estimators, the bias of all the estimators decrease as the sample size increases. When $\theta$  increases the estimates of the bias and mean squared error also increases for both estimators.  Additionally, the bias of the uncorrected estimators $\hat\pi$ is negative while that the bias of the corrected estimator $\hat\pi_{bc}$ is positive. About the mean squared error, we verify that it decreases as the sample size increases in all estimators, which is numerical indicative of the consistency of the estimators.

%
%For the evaluation of interval estimation we calculate the empirical coverage probability (ECP) and empirical tails coverage probability of each of the CI considered. In each Monte Carlo replication we calculated the confidence interval and checked if the CI contains the true parameter or not. The coverage rate is given by the percentage of replications in which the CI contained the parameter. It is desirable that the ECP value gets closer to the nominal coverage levels.
Table~\ref{ic} presents the results of numerical evaluation of interval estimators of the parameters of ZMPL distribution (for brevity, we only present results for $\theta=1.5$). For the parameter $\theta$, nominal coverages of 0.95 and 0.99, generally, the interval aCI had the best empirical coverages. Now for the parameter $\pi$, nominal coverages of 0.90 and 0.99, generally, the interval pCI had the best empirical coverages.

%\begin{landscape}
\begin{table}[t]
  \centering
\scriptsize
\caption{Empirical bias and mean squared errors (in parentheses).}
%\medskip
\resizebox{\linewidth}{!}{
\begin{tabular}{@{}cc  cc@{} c cc@{} c cc@{} c cc@{}}
\hline
           &       &\multicolumn{5}{c}{$\theta = 1.5$}&& \multicolumn{5}{c}{$\theta = 2.0$}\\ \cline{3-7} \cline{9-13} \\
   $n$ &$\pi$&   $\widehat{\theta}$&  $\widehat{\theta}_{bc}$ & &$\widehat{\pi}$&  $\widehat{\pi}_{bc}$ & &   $\widehat{\theta}$&  $\widehat{\theta}_{bc}$ & &$\widehat{\pi}$&  $\widehat{\pi}_{bc}$ \\ \\
\hline
       &  $-$0.10  &  0.163 & -0.029  &&  -0.080  &  0.075  &&  0.322  & -0.103  && -0.139 & 0.073  \\
       &        &(0.426)   &(0.216)   &&(0.119)  &(0.113)   &&(1.388) &(1.008)  &&(0.292)&(0.205)  \\
 35      &  0.00  &  0.198   &   -0.024    && -0.082  & 0.076 && 0.367  & -0.045   && -0.146 & 0.058  \\
       &        &(0.487)   &(0.219)  &&(0.114)  &(0.097) &&(1.527) &(1.228)  &&(0.275)&(0.199)  \\
   &  0.10  & 0.230    & -0.013   &&  -0.092   &  0.072  && 0.371  & -0.001 && -0.131  & 0.056  \\
       &        &(0.572)   &(0.271) &&(0.121)  &(0.104)  &&(1.511) &(1.244) &&(0.236)&(0.155)  \\ \\
%       &  0.30  & x.xxxx    & x.xxxx       &x.xxxx       &x.xxxx       && x.xxxx   & x.xxxx  &x.xxxx   & x.xxxx  && x.xxxx  & x.xxxx  &x.xxxx  & x.xxxx  && x.xxxx & x.xxxx&x.xxxx  & x.xxxx   \\
%       &        &(x.xxxx)   &(x.xxxx)      &(x.xxxx)       &x.xxxx       &&(x.xxxx)  &(x.xxxx) &x.xxxx   & x.xxxx  &&(x.xxxx) &(x.xxxx)&x.xxxx  & x.xxxx   &&(x.xxxx)&(x.xxxx)&x.xxxx  & x.xxxx  \\\\
       %
%       &  $-$0.30  & x.xxxx    & x.xxxx       &x.xxxx       &x.xxxx       && x.xxxx   & x.xxxx  &x.xxxx   & x.xxxx  && x.xxxx  & x.xxxx  &x.xxxx  & x.xxxx  && x.xxxx & x.xxxx&x.xxxx  & x.xxxx  \\
%       &        &(x.xxxx)   &(x.xxxx)      &x.xxxx       &x.xxxx       &&(x.xxxx)  &(x.xxxx) &x.xxxx   & x.xxxx  &&(x.xxxx) &(x.xxxx) &x.xxxx  & x.xxxx  &&(x.xxxx)&(x.xxxx)&x.xxxx  & x.xxxx  \\
       &  $-$0.10  & 0.090 & 0.005  && -0.042   & 0.018   && 0.193  & -0.042  && -0.080 & 0.027  \\
       &        &(0.162)   &(0.100) &&(0.052)  &(0.045) &&(0.690) &(0.553) &&(0.144)&(0.120)  \\
  60     &  0.00  & 0.096    &  -0.002 && -0.043  & 0.023   && 0.212  & -0.042 && -0.085 & 0.022 \\
       &        &(0.179)   &(0.107) &&(0.050)  &(0.046) &&(0.617) &(0.385) &&(0.107)&(0.070)  \\
   &  0.10  & 0.116    & 0.001   && -0.046   & 0.028  && 0.217  & -0.037  && -0.069 & 0.032  \\
       &        &(0.228)   &(0.134)  &&(0.051)  &(0.047)  &&(0.717) &(0.541)  &&(0.099)&(0.071)  \\ \\
%       &  0.30  & x.xxxx    & x.xxxx       &x.xxxx       &x.xxxx       && x.xxxx   & x.xxxx  &x.xxxx   & x.xxxx  && x.xxxx  & x.xxxx  &x.xxxx  & x.xxxx  && x.xxxx & x.xxxx&x.xxxx  & x.xxxx   \\
%       &        &(x.xxxx)   &(x.xxxx)      &x.xxxx       &x.xxxx       &&(x.xxxx)  &(x.xxxx) &x.xxxx   & x.xxxx  &&(x.xxxx) &(x.xxxx)&x.xxxx  & x.xxxx  &&(x.xxxx)&(x.xxxx)&x.xxxx  & x.xxxx  \\\\
       %
%           &  $-$0.30  & x.xxxx    & x.xxxx       &x.xxxx       &x.xxxx       && x.xxxx   & x.xxxx  &x.xxxx   & x.xxxx  && x.xxxx  & x.xxxx  &x.xxxx  & x.xxxx  && x.xxxx & x.xxxx&x.xxxx  & x.xxxx  \\
%       &        &(x.xxxx)   &(x.xxxx)      &x.xxxx       &x.xxxx       &&(x.xxxx)  &(x.xxxx) &x.xxxx   & x.xxxx  &&(x.xxxx) &(x.xxxx) &x.xxxx  & x.xxxx  &&(x.xxxx)&(x.xxxx)&x.xxxx  & x.xxxx  \\
       &  $-$0.10  &  0.056    & 0.010  && -0.029   & 0.003   && 0.120  & -0.018  && -0.048 & 0.017  \\
       &        &(0.091)   &(0.066)  &&(0.031)  &(0.028)  &&(0.289) &(0.201)  &&(0.064)&(0.051)  \\
  90     &  0.00  & 0.062    & 0.009  && -0.028   & 0.010   && 0.127  & -0.025   && -0.050 & 0.015  \\
       &        &(0.105)   &(0.076) &&(0.031)  &(0.029) &&(0.317) &(0.218)  &&(0.059)&(0.046) \\
    &  0.10  & 0.077    & 0.014   && -0.032    & 0.011   && 0.164  & -0.005   && -0.057 & 0.009  \\
       &        &(0.126)   &(0.088)  &&(0.029)  &(0.027)  &&(0.428) &(0.311)  &&(0.061)&(0.046)  \\ \\
%       &  0.30  & x.xxxx    & x.xxxx       &x.xxxx       &x.xxxx       && x.xxxx   & x.xxxx  &x.xxxx   & x.xxxx  && x.xxxx  & x.xxxx  &x.xxxx  & x.xxxx  && x.xxxx & x.xxxx&x.xxxx  & x.xxxx   \\
%       &        &(x.xxxx)   &(x.xxxx)      &x.xxxx       &x.xxxx       &&(x.xxxx)  &(x.xxxx) &x.xxxx   & x.xxxx  &&(x.xxxx) &(x.xxxx) &x.xxxx  & x.xxxx  &&(x.xxxx)&(x.xxxx)&x.xxxx  & x.xxxx  \\\\
       %
%           &  $-$0.30  & x.xxxx    & x.xxxx       &x.xxxx       &x.xxxx       && x.xxxx   & x.xxxx  &x.xxxx   & x.xxxx  && x.xxxx  & x.xxxx  &x.xxxx  & x.xxxx  && x.xxxx & x.xxxx&x.xxxx  & x.xxxx  \\
%       &        &(x.xxxx)   &(x.xxxx)      &x.xxxx       &x.xxxx       &&(x.xxxx)  &(x.xxxx) &x.xxxx   & x.xxxx  &&(x.xxxx) &(x.xxxx) &x.xxxx  & x.xxxx  &&(x.xxxx)&(x.xxxx)&x.xxxx  & x.xxxx  \\
       &  $-$0.10  & 0.045    & 0.018 && -0.023  & 0.000   && 0.090  & -0.004  && -0.039 & 0.006  \\
       &        &(0.066)   &(0.050) &&(0.023)  &(0.022) &&(0.195) &(0.152)  &&(0.044)&(0.037)  \\
 120      &  0.00  & 0.048   & 0.015  && -0.024  & 0.003  && 0.111 & 0.004  && -0.044  & 0.002  \\
       &        &(0.074)   &(0.056) &&(0.023)  &(0.022) &&(0.240)&(0.179)  &&(0.043) &(0.036)  \\
   &  0.10  & 0.057   & 0.019    && -0.021   & 0.013  && 0.106 & -0.013 && -0.038  & 0.008  \\
       &        &(0.087)   &(0.064)  &&(0.022)  &(0.023)  &&(0.250)&(0.187)   &&(0.037) &(0.031)  \\
%       &  0.30  & x.xxxx    & x.xxxx       &x.xxxx       &x.xxxx       && x.xxxx   & x.xxxx  &x.xxxx   & x.xxxx  && x.xxxx & x.xxxx &x.xxxx  & x.xxxx   && x.xxxx  & x.xxxx&x.xxxx  & x.xxxx  \\
%       &        &(x.xxxx)   &(x.xxxx)      &x.xxxx       &x.xxxx       &&(x.xxxx)  &(x.xxxx) &x.xxxx   & x.xxxx  &&(x.xxxx)&(x.xxxx)&x.xxxx  & x.xxxx   &&(x.xxxx) &(x.xxxx)&x.xxxx  & x.xxxx  \\
       %
\hline
\end{tabular}
}
\label{tabaplic}
\end{table}
%\end{landscape}

%\begin{landscape}
\begin{table}[t]
  \centering
\caption{Empirical coverage probability and empirical tails coverage probability (Left ; Right) of the confidence intervals for $\theta=1.5$ .}\label{ic}
%\medskip
\resizebox{\linewidth}{!}{
\begin{tabular}{@{}c@{}c@{}c@{}c@{}c@{} c@{} c@{} c@{}c@{}c@{}c@{} c@{} c@{}c@{}c@{}c@{} c@{} c@{}c@{}c@{}c@{}}
\hline
           &       &\multicolumn{9}{c}{$\theta$}&& \multicolumn{9}{c}{$\pi$}\\ \cline{3-11} \cline{13-21} \\
   $n$ &$\pi$&   $\text{aCI}(\cdot;0.90)$ &    $\text{pCI}(\cdot;0.90)$  &   $\text{aCI}(\cdot;0.95)$ & $\text{pCI}(\cdot;0.95)$ & & $\text{aCI}(\cdot;0.99)$ &    $\text{pCI}(\cdot;0.99)$  &   &  & &   $\text{aCI}(\cdot;0.90)$&  $\text{pCI}(\cdot;0.90)$&  $\text{aCI}(\cdot;0.95)$&  $\text{pCI}(\cdot;0.95)$& &$\text{aCI}(\cdot;0.99)$& $\text{pCI}(\cdot;0.99)$& &  \\ \\
\hline
       &  $-$0.10  &   0.925   &   \bf{0.891}     &  \bf{0.949}      &  0.939   && \bf{0.980}  & 0.976  &   &   &&  0.924     & \bf{0.916} & \bf{0.955} &  0.959 && 0.985 &\bf{0.988} &  &  \\
       &        &(0.001;0.074)   &(0.089;0.020)      &(0.000;0.050) & (0.054;0.007)       &&(0.000;0.020)  &(0.024;0.000) &   &   &&(0.065;0.010) &(0.005;0.080)&(0.042;0.003)  & (0.000;0.041)   &&(0.015;0.000)&(0.000;0.012)& &   \\

  35     &  0.00  &  0.929   &   \bf{0.881}     &  \bf{0.955}     &   0.932    && \bf{0.983}    &  0.976 &   &   &&  0.919 &  \bf{0.910}   &\bf{0.955}  &  0.956 && 0.982 &\bf{0.988}  & &   \\
       &        &(0.000;0.071)   &(0.102;0.016)      &(0.000;0.045)       &(0.064;0.004) &&(0.000;0.017)  &(0.024;0.000) &   &   &&(0.072;0.010) &(0.007;0.082) &(0.044;0.002)  & (0.000;0.044)  &&(0.018;0.000)&(0.000;0.012)&  &   \\

     &  0.10  & 0.929    & \bf{0.886}      &\bf{0.953}       &0.933       && \bf{0.982}   & 0.974  &   &   && 0.923 & \bf{0.910}  &0.955  & \bf{0.954} && 0.985 & \bf{0.987}&  &  \\
       &        &(0.000;0.071)   &(0.101;0.013)      &(0.000;0.047)       &(0.064;0.004)       &&(0.000;0.018)  &(0.026;0.000) &   &   &&(0.067;0.010) &(0.009;0.081) &(0.043;0.002)  & (0.001;0.045)  &&(0.015;0.000)&(0.000;0.013)&  &    \\\\
     &  $-$0.10  &  0.924   & \bf{0.892}       &\bf{0.957}       &0.942      && \bf{0.987}   & 0.983  &   &   &&  0.919  & \bf{0.910}  &0.960  & \bf{0.959}  && 0.988 & \bf{0.991}&  &  \\
       &        &(0.01;0.066)   &(0.084;0.024)      &(0.000;0.043) & (0.051;0.006)       &&(0.000;0.013)  &(0.017;0.000) &   &   &&(0.061;0.019) &(0.023;0.067)&(0.037;0.003)  & (0.005;0.036)   &&(0.012;0.000)&(0.000;0.009)& &   \\

  60     & 0.00  & 0.925    & \bf{0.900}       &0.958       &\bf{0.945}       && \bf{0.986}   & 0.982  &   &   && 0.917  & \bf{0.904}  &0.956  & 0.956  && 0.988 & \bf{0.990}&   \\
       &        &(0.007;0.068)   &(0.079;0.021)      &(0.000;0.042)       &(0.047;0.009) &&(0.000;0.014)  &(0.017;0.001) &   &   &&(0.063;0.020) &(0.022;0.073) &(0.040;0.004)  & (0.004;0.040)  &&(0.012;0.000)&(0.000;0.010)&  &   \\

     &  0.10  & 0.922    & \bf{0.896}      &\bf{0.953}       &0.942       && \bf{0.984}   & 0.982  &   &   && 0.912 & \bf{0.906}  &\bf{0.955}  & 0.960  && 0.983 & \bf{0.989}&  &  \\
       &        &(0.005;0.073)   &(0.081;0.023)      &(0.000;0.047)       &(0.051;0.006)       &&(0.000;0.016)  &(0.018;0.000) &   &   &&(0.069;0.019) &(0.022;0.072) &(0.041;0.004)  & (0.002;0.038)  &&(0.016;0.000)&(0.000;0.011)&  &    \\\\
        &  $-$0.10  & 0.915    & \bf{0.896}       &0.956       &\bf{0.948}      && \bf{0.986}   & 0.985  &   &   &&     0.917  & \bf{0.901}  &0.958  & \bf{0.952}  && 0.987 & \bf{0.992}&  &  \\
       &        &(0.017;0.068)   &(0.075;0.029)      &(0.003;0.041) & (0.040;0.012)       &&(0.000;0.014)  &(0.014;0.001) &   &   &&(0.058;0.025) &(0.033;0.067)&(0.035;0.007)  & (0.014;0.035)   &&(0.012 ;0.000)&(0.000;0.008)& &   \\

  90     & 0.00  & 0.913    & \bf{0.893}       &0.958       &\bf{0.943}       && \bf{0.986}   & 0.982  &   &   && 0.908  & \bf{0.898}  &0.957  & \bf{0.955}  && 0.988 & \bf{0.991} &   \\
       &        &(0.020;0.067)   &(0.077;0.030)      &(0.002;0.040)       &(0.046;0.011) &&(0.000;0.014)  &(0.017;0.001) &   &   &&(0.064;0.028) &(0.029;0.073) &(0.036;0.007)  & (0.007;0.038)  &&(0.012;0.000)&(0.000;0.009)&  &   \\

     &  0.10  & 0.916    & \bf{0.896}      &\bf{0.957}       &0.939       && \bf{0.987}   & 0.983  &   &   && 0.915 & \bf{0.910}  &\bf{0.959}  & 0.961  && 0.987 &\bf{0.990}&  &  \\
       &        &(0.018;0.066)   &(0.078;0.026)      &(0.003;0.040)       &(0.050;0.011)       &&(0.000;0.013)  &(0.016;0.001) &   &   &&(0.058;0.027) &(0.021;0.068) &(0.034;0.008)  & (0.002;0.036)  &&(0.012;0.000)&(0.000;0.010)&  &    \\\\
     &  $-$0.10  & 0.908    & \bf{0.895}       &\bf{0.953}       &0.943      && \bf{0.987}   & 0.986  &   &   &&     0.910  & \bf{0.903}  &\bf{0.953}  & 0.958  && 0.989 & \bf{0.990}&  &  \\
       &        &(0.027;0.064)   &(0.073;0.031)      &(0.007;0.040) & (0.043;0.014)       &&(0.000;0.013)  &(0.013;0.002) &   &   &&(0.057;0.033) &(0.031;0.066)&(0.036;0.012)  & (0.007;0.035)   &&(0.011;0.000)&(0.000;0.010)& &   \\

  120     & 0.00  & 0.909    & \bf{0.893}       &\bf{0.954}       &0.944       && \bf{0.987}   & 0.986  &   &   && \bf{0.903}  & 0.913  &\bf{0.952}  & 0.962  && 0.989 & \bf{0.991} &   \\
       &        &(0.024;0.067)   &(0.075;0.032)      &(0.007;0.039)       &(0.042;0.014) &&(0.000;0.013)  &(0.013;0.001) &   &   &&(0.063;0.034) &(0.020;0.067) &(0.036;0.012)  & (0.002;0.037)  &&(0.010;0.001)&(0.000;0.009)&  &   \\

     &  0.10  & \bf{0.907}    & 0.891      &\bf{0.954}       &0.943       && \bf{0.986}   & 0.984  &   &   && \bf{0.905} & 0.918  &\bf{0.948}  & 0.962  && 0.984 & \bf{0.990}&  &  \\
       &        &(0.023;0.070)   &(0.076;0.032)      &(0.005;0.041)       &(0.044;0.013)       &&(0.000;0.014)  &(0.014;0.002) &   &   &&( 0.063;0.032) &(0.018;0.064) &(0.041;0.011)  & (0.003;0.036)  &&(0.016;0.001)&(0.000;0.010)&  &    \\\\
\hline
\end{tabular}
}
\end{table}
%\end{landscape}

%
%  120     & 0.00  & xxxxx    & xxxxx       &xxxxx       &xxxxx       && xxxxx   & xxxxx  &   &   && xxxxx  & xxxxx  &xxxxx  & xxxxx  && xxxxx & xxxxx &   \\
%       &        &(xxxxx)   &(xxxxx)      &(xxxxx)       &(xxxxx) &&(xxxxx)  &(xxxxx) &   &   &&(xxxxx) &(xxxxx) &(xxxxx)  & (xxxxx)  &&(xxxxx)&(xxxxx)&  &   \\
%
%     &  0.10  & xxxxx    & xxxxx      &x.xxxx       &x.xxxx       && xxxxx   & xxxxx  &   &   && xxxxx & xxxxx  &x.xxxx  & x.xxxx  && xxxxx & xxxxx&  &  \\
%       &        &(xxxxx)   &(xxxxx)      &(x.xxxx)       &(x.xxxx)       &&(xxxxx)  &(xxxxx) &   &   &&(xxxxx) &(xxxxx) &(x.xxxx)  & (x.xxxx)  &&(xxxxx)&(xxxxxx)&  &    \\\\

\section{Applications}\label{sec5}

In this section we apply the proposed methodology to two real demand data sets. Here, we perform an exploratory data analysis and, based on it, we show the good fitting of the ZMPL distribution to the analyzed data. As the samples are large, we estimate the unknown parameters of the fitted models by the ML method (as discussed
in Section 3).

\subsection{Example 1: Inflation of zeros}

In this section we have tried to fit Poisson distribution, Poisson-Lindley distribution and Zero-Modified Poisson-Lindley distribution to a biological data using maximum likelihood estimates. Here we use a dataset related to mammalian cytogenetic dosimetry lesions in rabbit lymphoblast induced by streptonigrin (NSC-45383), exposure - $60 \, \mu g\,|\,Kg$. For more details about  the data see \cite{sha15}.  Table~\ref{descriptive2} presents some descriptive measures to the data set studied in this application.  The skewness here is 2.42. This value implies that the distribution of the data is positively skewed. For the kurtosis, we have 10.70 implying that the distribution of the data is leptokurtic. The dispersion index shows that the data must be modeled by a overdispersed model (FI $= 1.56$).
The proportion of zeros in the data set is $69\%$. Then, we have evidence that there is inflation of zeros. Thus, the use of the ZMPL
distribution for fitting this data set appears justified.

% latex table generated in R 3.3.3 by xtable 1.8-2 package
% Thu May 18 13:03:19 2017
\begin{table}[H]
\centering
\caption{Descriptive Measures}
\resizebox{\linewidth}{!}{
\begin{tabular}{rrrrrrrr}
  \hline
Measures & Minimum & Maximum& Mean & Variance & Skewness & Kurtosis & FI \\
  \hline
Values & 0.00 & 6.00 & 0.47 &  0.74 & 2.42 & 10.70 & 1.56 \\
   \hline
\end{tabular}
}
\label{descriptive2}
\end{table}

Table \ref{tab5} provides the estimates of the model parameters and chi-squared test. The gradient statistic for testing $\mathcal{H}_0: \pi = 0$ is $S_g=114.49$ and $p\mbox{-value} <0.001$, i.e, the parameter $\pi$ is statistically  different from zero.

\begin{table}[H]
  \centering
    \small
\caption{Distribution of mammalian cytogenetic dosimetry lesions in rabbit lymphoblast induced by streptonigrin (NSC-45383), Exposure - $60 \, \mu g\,|\,Kg$ with expected frequency obtained by fitting Poisson, Poisson-Lindley and Zero-modified Poisson-Lindley distributions.}
%\medskip
\resizebox{\linewidth}{!}{
\begin{tabular}{cccccc}
\hline
\multirow{3}*{\parbox{3cm}{\centering N$^o$ of mammalian cytogenetic dosimetry lesions} }&\multirow{3}*{\parbox{2cm}{ \centering Observed frequency}}&\multicolumn{4}{c}{Expected frequency}\\ \cline{3-6}
                                                          &                                                                               &   \multirow{2}*{Poisson}& Zero-modified &  \multirow{2}*{Poisson-Lindley} & Zero-modified   \\
                                                          &                                                                               &          &       Poisson  &         & Poisson-Lindley       \\
\hline
0       &  413  &  374.0  &413.0 &   405.7      & 413.0   \\
1       &  124  &  177.4  &116.0 &   133.6      & 123.4   \\
2       &   42  &   42.1  &52.1 &    42.6      & 42.9   \\
3       &   15  &    6.6  &15.6 &    13.3      & 14.5   \\
4       &    5  &    0.8  &3.5 &     4.1      & 4.8   \\
5       &    0  &    0.1  &0.6 &     1.2      & 1.6    \\
6       &    2  &    0.0  &0.2 &     0.5      & 0.5   \\
\hline
Total        & 601   &  601 &601  &   601      & 601    \\ \hline
Estimate of  &    &   \multirow{2}*{$\hat \theta = 0.47421$}& $\hat \theta=0.8989$  & \multirow{2}*{$\hat \theta = 2.6854$}        & $\hat \theta=2.4098$     \\
parameter    &     &   & $\hat \pi=0.4725$  &         & $\hat \pi=0.1165$   \\ \hline
Confidence  &    &   \multirow{2}*{$\textrm{aCI}(\theta,0.95) = (0.4192;0.5293)$}& $\textrm{aCI}(\theta,0.95)=(0.7304;1.0675)$  & \multirow{2}*{$\textrm{aCI}(\theta,0.95)= (2.3619;3.0088)$}        & $\textrm{aCI}(\theta,0.95)=(1.8904;2.9290)$     \\
Interval    &     &   & $\textrm{aCI}(\pi,0.95)=(0.3852;0.5599)$  &         & $\textrm{aCI}(\pi,0.95)=(-0.0649;0.2979)$   \\ \hline
$\chi^2$     &    &   726.2816& 42.19  & 9.6880        & 5.9064   \\ \hline
d.f          &    &   6 & 6&     6   & 6   \\ \hline
$p$-value    &    & $<0.0001$  & $<0.0001$ & 0.1384        & 0.4338   \\ \hline
\end{tabular}
}
\label{tab5}
\end{table}

The standardized differences (SD) are calculated of the following form
$$
\Delta_{s d_i } = \frac{\delta_{s d_i}}{\max\left\{ \delta_{(\cdot) d_1}, \ldots, \delta_{(\cdot) d_m} \right\} }, \quad i=1,\ldots, m; s \in \mathcal{S},
$$
where $\delta_{s d_i} = (\text{observed}_{s d_i} - \text{expected}_{s d_i})$, $\delta_{(\cdot) d_i} = (\delta_{s_1 d_i}, \ldots, \delta_{s_{|\mathcal{S}|} d_i})$, $m$ is the number of studied models, $\mathcal{S}$ is the support (common) of the distributions and $|A|$ is the cardinality of a set $A$. The SD plots are formed by standardized versus common distributions of distributions. We do the this when we are interested in to know what model that best fits the data.

\begin{figure}[H]
	\centering
	\includegraphics[width=7.5cm,height=7.5cm]{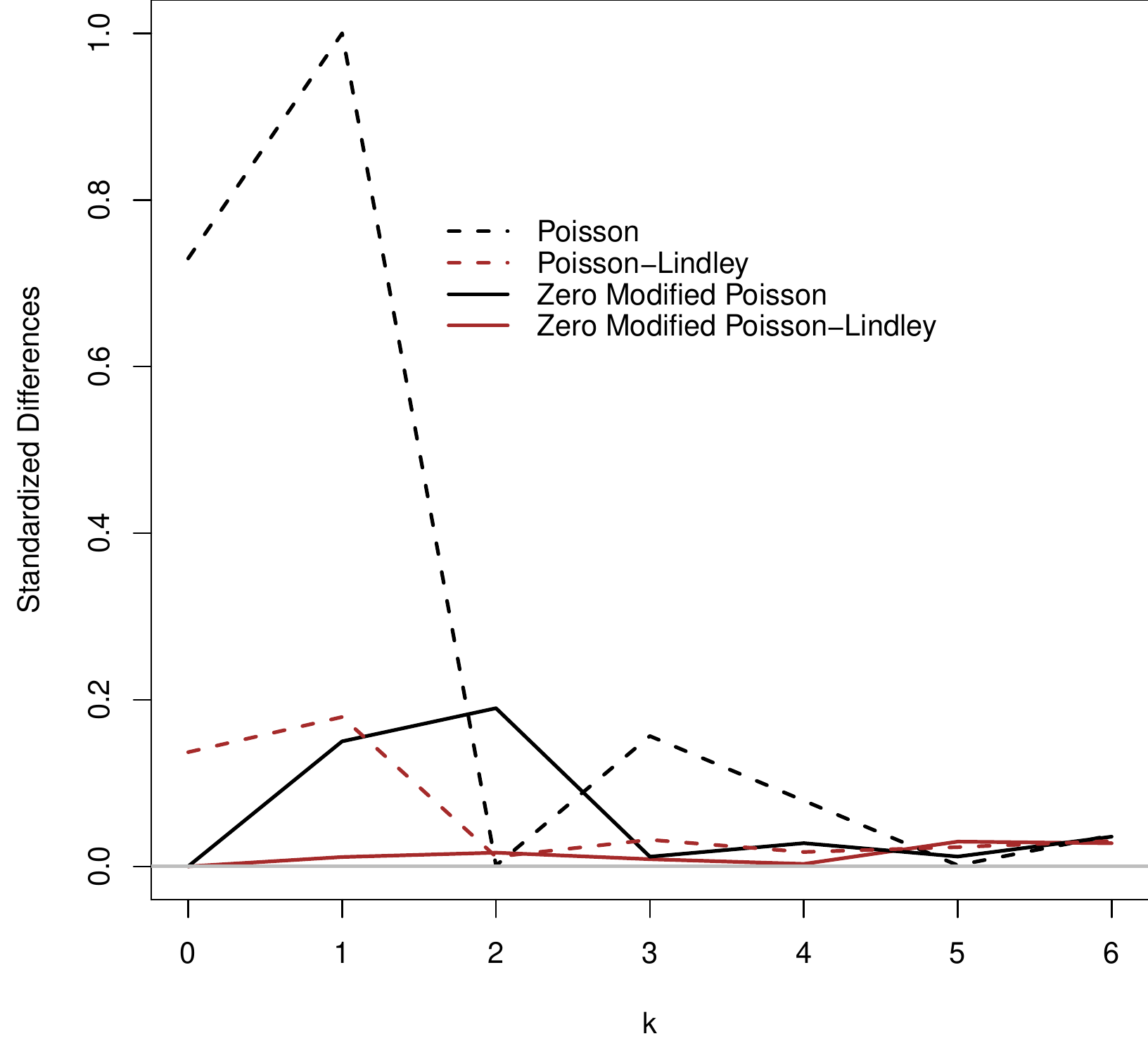}
	\caption{SD-plot for \cite{sha15}  data.}
	\label{figaplic2}
\end{figure}

The SD-plot of the fitted ZMPL, PL, ZMP and Poisson models are shown in Figure~\ref{figaplic2}. Based on the values of the Table~\ref{tab5}, we conclude that the ZMPL distribution provides a better fit than the Poisson, ZMP and PL distributions.

\subsection{Example 2: Deflation of zeros}

This dataset gives the number of outbreaks of strikes in the UK coal mining industry in sucessive four-week periods, in the years 1948-1959 \citep[][]{Ridout:2004aa}. These data are only modestly underdispersed, with FI of 0.75 (see, Table~\ref{descriptive3}).
The proportion of zeros in the data set is $29\%$. Then, we have evidence that
there is deflation of zeros. Thus, the use of the ZMPL
distribution for fitting this data set appears justified.

% latex table generated in R 3.3.3 by xtable 1.8-2 package
% Thu May 18 13:03:19 2017
\begin{table}[H]
\centering
\caption{Descriptive Measures}
\begin{tabular}{rrrrrrrr}
  \hline
Measures & Minimum & Maximum& Mean & Variance & Skewness & Kurtosis & FI \\
  \hline
Values & 0.00 & 4.00 & 0.99 &  0.74 & 0.80 & 3.5 & 0.75 \\
   \hline
\end{tabular}
\label{descriptive3}
\end{table}

Table \ref{tab6} provides the estimates of the model parameters and chi-squared test. The gradient statistic for testing $\mathcal{H}_0: \pi = 0$ is $S_g=5275.1$ and $p\mbox{-value} <0.001$, i.e, the parameter $\pi$ is statistically  different from zero.

\begin{figure}[H]
\centering
\includegraphics[width=7.5cm,height=7.5cm]{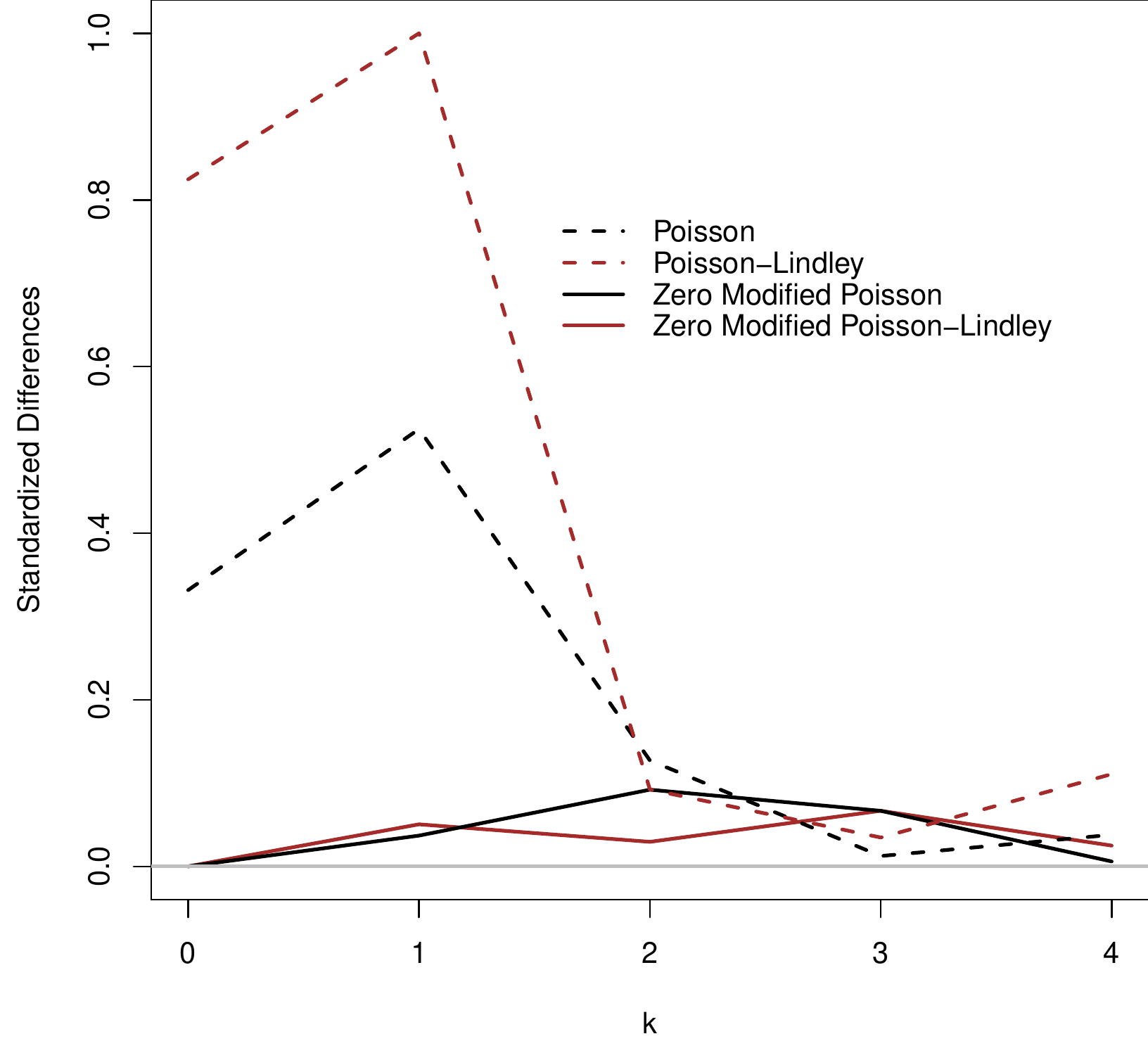}
\caption{SD-plot for \cite{Ridout:2004aa}  data.}
\label{figaplic3}
\end{figure}

Figure~\ref{figaplic3} presents the SD-plot for the fitted models.
Based on the values of the Table~\ref{tab6} and Figure~\ref{figaplic3},  we observe that the ZMP and ZMPL models are competitive.

\begin{table}[H]
  \centering
    \small
\caption{Fitted frequencies, estimate of parameters and chi-square statistics from fitted four distributions to the strike outbreak data of \cite{Ridout:2004aa}.}
%\medskip
\resizebox{\linewidth}{!}{
\begin{tabular}{cccccc}
\hline
\multirow{3}*{\parbox{3cm}{\centering N$^o$ of outbreaks} }&\multirow{3}*{\parbox{2cm}{ \centering Observed frequency}}&\multicolumn{4}{c}{Expected frequency}\\ \cline{3-6}
                                                          &                                                                               &   \multirow{2}*{Poisson}& Zero-modified &  \multirow{2}*{Poisson-Lindley} & Zero-modified   \\
                                                          &                                                                               &          &       Poisson  &         & Poisson-Lindley       \\
\hline
0          &  46   &  57.76  &46.00  &75.24    & 46.00   \\
1           &  76  & 57.39   &74.69  & 40.55   &77.79    \\
2           &  24  &  28.51  &27.27  & 20.73   & 22.95   \\
3           &   9   &  9.44  &6.64  & 10.23   & 6.63   \\
$\geq$4& 1     &  2.35  &1.21  &  4.93   & 1.89    \\
\hline
Total        & 156   &  156 &156  &   156      & 156    \\ \hline
Estimate of  &    &   \multirow{2}*{$\hat \theta = 0.9936$}& $\hat \theta=0.7301$  & \multirow{2}*{$\hat \theta = 1.4010$}        & $\hat \theta=2.9579$    \\
parameter    &     &   & $\hat \pi=-0.3609$  &         & $\hat \pi=-1.3475$   \\ \hline
Confidence  &    &   \multirow{2}*{$\textrm{aCI}(\theta,0.95) = (0.8372;1.1500)$}& $\textrm{aCI}(\theta,0.95)=(0.5271;0.9331)$  & \multirow{2}*{$\textrm{aCI}(\theta,0.95)= (1.1478;1.6542)$}        & $\textrm{aCI}(\theta,0.95)=(2.0436;3.8721)$     \\
Interval    &     &   & $\textrm{aCI}(\pi,0.95)=(-0.6526;-0.0691)$  &         & $\textrm{aCI}(\pi,0.95)=(-1.9923;-0.7028)$   \\ \hline
$\chi^2$     &    &   9.8986& 1.2916  & 44.748       &1.3404  \\ \hline
d.f          &    &   4 & 4 &     4   & 4   \\ \hline
$p$-value    &    & 0.04217  & 0.8628 & $<0.0001$       & 0.8545  \\ \hline
\end{tabular}
}
\label{tab6}
\end{table}

\section{Concluding remarks}\label{sec6}

The paper has introduced the zero-modified Poisson-Lindley distribution. The paper has derived maximum likelihood estimators and of their corrected version. Two applications of the ZMPL distribution illustrated the usefulness of the model. A main theme of the paper has been the comparison of the efficiency of the ML and of their corrected version. It is shown that the biases of the bias-corrected estimators have good finite-sample behavior, outperforming the ML estimator. Two different strategies for interval estimation were considered and numerically evaluated. Thus, zero-modified distributions are good candidates when excess of zeros appears.  Zero-modified distributions should be the solution of this problem to avoid any over or under estimation of the measure of interest.  In conclusion, it is believed that the ZMPL distribution and subsequent regression model may offer a very useful tool for analyzing data characterized with a large or small amount of zeros.
%
%rPL <- function(n, theta)
%{
%    x <- rep(NA, times = n)
%    p <- theta/(theta + 1)
%	for(i in 1:n){
%		u  <- runif(1)
%		y1 <- rgamma(1, 1, rate = theta)
%		y2 <- rgamma(1, 1, rate = theta)
%		if(u <= p){
%			x[i] <- rpois(1, y1)
%		}
%		else{
%			x[i] <- rpois(1, y1 + y2)
%		}
%	}
%return(x)
%}
%\section*{Acknowledgements} %Colocar os agradecimentos na última versão do artigo
%We gratefully acknowledge grants from CAPES and CNPq (Brazil). Research developed with the computational resources of the Center for Mathematical Sciences Applied to Industry (CeMEAI) financed by FAPESP.

%\section*{References}
%\bibliographystyle{authordate1}
%%\scriptsize
%\bibliography{ZMPL}

%\section*{References}
%\begin{thebibliography}{99}
%

%
%\bibitem[Ghitany and Al-Mutairi(2009)]{ghi09}
%Ghitany, M.E. and Al-Mutairi, D. K. (2009). Estimation Methods for the discrete Poisson-Lindley distribution. \emph{Journal of Statistical Computation and Simulation}, {\bf 79}, 1--9.
%
%\bibitem[Ghitany et al.(2008)]{ghi08}
%Ghitany, M. E., Al-Mutairi, D. K. and Nadarajah, S. (2008). Zero-truncated Poisson-Lindley distribution and its application. \emph{Mathematics and Computers in Simulation}, {\bf 79}, 279--287.
%
%\bibitem[Sankaran(1970)]{san70}
%Sankaran, M. (1970). The discrete Poisson-Lindley distribution. \emph{Biometrics}, {\bf 26}, 145--149.
%
%
%\bibitem[Shanker and Fesshaye(2015)]{sha15}
%Shanker, R. and Fesshaye, H. (2015). On Poisson-Lindley Distribution and Its Applications to Biological Sciences. \emph{Biometrics \& Biostatistics International Journal}, {\bf 2}, 1--5.
%
%
%\end{thebibliography}

\end{document}